# Fully Automated Multi-Organ Segmentation in Abdominal Magnetic Resonance Imaging with Deep Neural Networks

Yuhua Chen[1,2], Dan Ruan[1,3], Jiayu Xiao[2], Lixia Wang[2,4], Bin Sun[5], Rola Saouaf[6], Wensha Yang[7], Debiao Li[1,2,8], Zhaoyang Fan[1,2,8]

1. Department of Bioengineering, University of California, Los Angeles, CA, USA;
2. Biomedical Imaging Research Institute, Cedars-Sinai Medical Center, Los Angeles, CA, USA;
3. Department of Radiation Oncology, University of California, Los Angeles, CA, USA;
4. Department of Radiology, Chaoyang Hospital, Capital Medical University, Beijing, China;
5. Department of Radiology, Fujian Medical University Union Hospital, Fuzhou, Fujian, China;
6. Department of Imaging, Cedars-Sinai Medical Center, Los Angeles, CA, USA;
7. Department of Radiation Oncology, University of Southern California, Los Angeles, CA, USA;
8. Department of Medicine, University of California, Los Angeles, CA, USA.

# ABSTRACT

**Purpose**

Segmentation of multiple organs-at-risk (OARs) is essential for radiation therapy treatment planning and other clinical applications. Current practice requires manual delineation, which is time-consuming, labor-intensive, and prone to intra- and inter-observer variations. We developed a deep learning (DL) framework for fully automated segmentation of multiple OARs in clinical abdominal MR imaging with high accuracy, reliability, and efficiency.

**Methods**

We developed an Automated deep Learning-based Abdominal Multi-Organ segmentation (ALAMO) framework based on 2D U-net and a densely connected network structure with tailored design in data augmentation and training procedures such as deep connection, auxiliary supervision, and multi-view. The model takes in multi-slice MR images and generates the output of segmentation results. Three-Tesla T1 VIBE (Volumetric Interpolated Breath-hold Examination) images of 102 subjects were collected and used in our study. Ten OARs were studied, including the liver, spleen, pancreas, left/right kidneys, stomach, duodenum, small intestine, spinal cord, and vertebral bodies. Two radiologists manually labeled and obtained the consensus contours as the ground-truth.  In the complete cohort of 102, 20 samples were held out for independent testing, and the rest were used for training and validation. The performance was measured using volume overlapping and surface distance.

**Results**

The ALAMO framework generated segmentation labels in good agreement with the manual results. Specifically, among the 10 OARs, 9 achieved high Dice Similarity Coefficients (DSCs) in the range of 0.87-0.96, except for the duodenum with a DSC of 0.80. The inference completes within one minute for a 3D volume of 320x288x180. Overall, the ALAMO model matches the state-of-the-art performance.

**Conclusion**

The proposed ALAMO framework allows for fully automated abdominal MR segmentation with high accuracy and low memory and computation time demands.

# 1. INTRODUCTION

The location and structural information of internal organs are essential in several clinical applications such as radiation therapy (RT), imaging-guided surgery, lesion quantification [1]. Such information is obtained routinely by human annotations on medical images, which is labor-intensive, time-consuming, and prone to intra- and inter-observer inconsistency. These limitations have also hindered certain timing-sensitive applications such as adaptive abdominal RT. Automated multi-organ segmentation is a compelling solution but remains a critical challenge because of complicated internal structures and widely variable organ sizes [2].

Early research on automated segmentation algorithms focused on mathematical modeling of the morphological information of organs. For instance, level-set [3], SNAKE [4], and graph cut [5] focus on attracting descriptors to organ boundaries, driven by intensity gradient and neighborhood structures. However, these models usually rely on the consistent appearance of edges and intensity patterns as well as specific scale tradeoffs, which limits their applicability to magnetic resonance (MR) datasets that commonly exhibit heterogeneous image quality. MR images are affected by variations in many factors, such as system models and manufacturers, sequence parameter settings, and field shimming conditions. Data-driven methods, such as atlas-based approaches [6], were investigated as another solution. The major drawbacks of atlas-based methods are their heavy dependence on atlas quality and size, and the consistency requirement between the target and atlas samples. Besides, the long processing time needed for performing multiple registrations poses the main hurdle for its practical use.

Recently, learning-based approaches [1] that combine modeling and data-driven methods became popular. Unlike previous methods with manually crafted morphological features, learning-based methods, specifically deep neural networks, learn the representative features directly from training data [7]. Its superior ability to model the complexity in multi-organ shape, context information, and the variety from inter-subject difference has been demonstrated on several benchmark datasets [8]. MR image-based segmentation tasks for the brain [9], heart [10], and breast [11] have been studied using deep learning (DL) and artificial intelligence techniques. However, there are very few studies focused on abdominal MR [12-15]. Despite substantial improvement over the years, the performance in automated abdominal MR segmentation still does not match up to the human performance, particularly in complex-structure organs such as the stomach and duodenum [14]. Most of the previous studies utilized 2D neural networks for organ segmentation were from generic computer vision. However, single-channel 2D models are insufficient to analyze 3D complex structures in volumetric medical images and motivates our investigation of a multi-slice setting. Another challenge for the application of DL in abdominal MR is overfitting, caused by the small data size – one reason that computed tomography (CT) based studies are more common and successful than MR counterparts. With sample size in the order of hundreds or less, increasing the network complexity does not lead to performance gain. We conjecture that a carefully designed network structure with more effective use of the existing nodes such as skip connections [16] is likely to be more beneficial in expanding the representation power without risking overfitting. Finally, the small training size also necessitates improvement in the training procedures, utilizing data augmentation, and deeply supervised learning [17].

In this work, we proposed a convolutional neural network (CNN) based fully automated MR image-based multi-organ segmentation technique, namely *ALAMO* (**A**utomated deep **L**earning-based **A**bdominal **M**ulti-**O**rgan segmentation). A multi-slice 2D neural network was developed to account for the correlative as well as complementary information between adjacent slices in the intrinsic 3D space while avoiding the heavy computation burden. To improve robustness and reduce overfitting risk, we investigated multiple approaches, including network normalization, data augmentation, and deeply supervised learning. We also introduced a novel multi-view training and inference technique that is simple, fast, yet effective to remove outliers in the preliminary segmentation predictions.

## 2. METHODS

### 2.1. MR Data

This work was built on a routinely used 3D abdominal MR sequence - T1-VIBE (Volumetric Interpolated Breath-hold Examination) [18]. Images were acquired from multiple 3-Tesla systems of the same manufacturer (Siemens Healthineers). A total of 102 subjects with no significant lesion presence (no or <2cm) were retrospectively enrolled in this study, with Institutional Review Board approval. We further split those cases into 66 for training, 16 for validation, and 20 for testing. Each T1-VIBE dataset consisted of 72 - 80 transversal slices with a spatial resolution of 1.1-1.3 mm in each 2D transversal slice and 2.0-4.0 mm in slice thickness. All 3D image sets were interpolated into 1.2 mm isotropic resolution. Two experienced radiologists independently labeled 10 organs (liver, spleen, pancreas, left/right kidneys, stomach, duodenum, small intestine, spinal cord, and vertebral bodies) and then reached a consensus contour.

### 2.2. Data Preprocessing and Augmentation

Dataset standardization, a common practice to speed up DL model training and improve network performance [19], was performed. Specifically, the mean signal intensity of each 3D dataset was subtracted from each voxel. The resultant voxel signal intensity was then normalized by the standard deviation of signal intensities in each 3D dataset.

During the training, 20 contiguous transversal slices in a matrix size of 256x160 were randomly cropped from the whole 3D volume which was typically 320x288x180. Random up-down or left-right flipping was applied at a probability of 50% for data augmentation. We further investigated the effect of random projective deformation as extra steps for data augmentation. Specifically, the image was transformed by a projection matrix (rotation angle: -0.05 -- + 0.05 rad, shearing scale: -0.3 -- +0.3, projective scale: -0.003 -- + 0.003) at a probability of 50%.

### 2.3. Deep Learning Models and Framework

To capture both high-resolution local textures and low-resolution context information, the ALAMO framework adopts the renowned U-net structure [20, 21]. The encoder of the network takes the input images on a full resolution scale and then gradually reduces the size of the feature maps to abstract the context information. The decoder of the network then takes in encoded features and provides the annotation prediction in a coarse-to-fine manner. At each resolution, a couple of convolutional layers are used in a block to process image features. In our work, the blocks were interchangeable between plainly-stacked layers and densely-connected layers, corresponding to PlainUnet and DenseUnet, respectively. Studies [16, 22, 23] have shown that DenseNet is less prone to overfitting problems. Hence, DenseUnet was optimized and evaluated against PlainUnet in this work. In DenseUnet, the growth-rate $k$ defines the number of feature maps in a single convolutional layer. In PlainUnet, the initial feature map number $f$ is the number of filters in the first resolution level and doubles for each lower resolution [20]. The ALAMO structure is illustrated in Figure. 1.

To fully capture the 3D information in a 2D network, we introduced a multi-slice input with a multi-channel 2D network structure. The 20 contiguous slices obtained from the whole 3D volume serve as the network input. The first layer of the network is a multi-channel convolutional layer. The multi-slice mechanism enables us to feed information about the third dimension to the network even though the network is 2D operational.

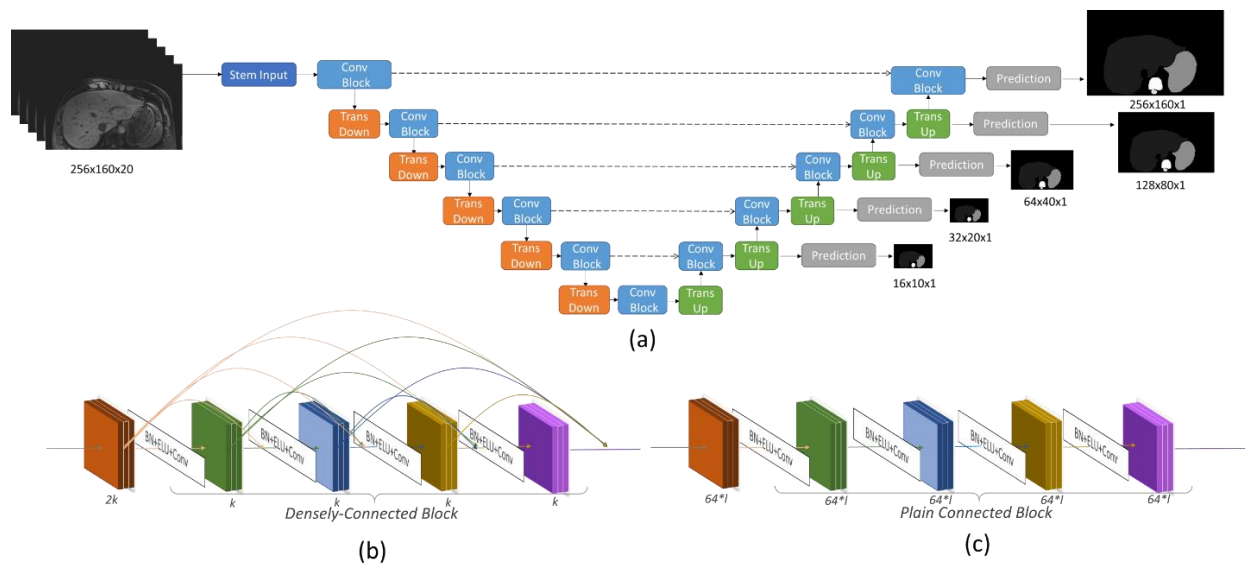

**Figure 1.** Network structure: (a) The U-net structure, the convolutional blocks are either the (b) Densely-Connected Block or (c) Plain Connected Block. The stem input module is added after the input. It contains a single 3x3 convolutional layer with the filter number of *f* or 2*k*. The kernel size is 3x3. Normalization is applied before the Exponential Linear Units (ELU) activation. 2x2 average pooling is used in the Transition Down, and a transposed-convolutional layer with 2x2 kernel and stride size of 2 is applied in the Transition Up. In the PlainUnet setting, the filter number starts at *f*=64 and doubles after pooling; in DenseUnet, the growth-rate *k*=48. A final layer with 1x1 kernel and softmax activation output the predictions for 11 classes (background + 10 organs). At each resolution, we use the auxiliary prediction for deeply supervised training.

### 2.4. Network Normalization

Network normalization is a widely used approach to help accelerate convergence, stabilize gradients, and alleviate overfitting to training data. Specific implementation options include batch normalization (BN) [14, 21, 23, 24], Instance normalization (IN) [25] and layer normalization (LN) [26]. However, a recent study shows that the batch size heavily affects the normalization performance [24]. Due to the memory limitation, we can only fit a single image per batch. We expect that normalization would have an incidental role in this specific scenario, with simple-sample driven statistics. For the sake of being comprehensive, we investigated the effect of using BN, using BN with the training model during testing [21], as well as IN and LN.

### 2.5. Deeply Supervised Training

One of the challenges in training a deep neural network is that the gradient tends to vanish in the backpropagation process. The skip connections in DenseUnet help to partially alleviate the problem in the same resolution level. Nevertheless, during the scaling process across different resolutions, it would be beneficial to direct gradients for optimizing the layers at a granular level. Deeply supervised training [17] is an approach to add auxiliary side predictions to down-sampled labels directly from the output of each resolution. Therefore, we added extra layers for low-resolution predictions in the decoder branches, as shown in Figure 1. The network is optimized to produce not only full-resolution segmentation masks but also multiple low-resolution masks. During the testing phase, the model only computes the final full-resolution output. Therefore, the low-resolution auxiliary predictions would not add any computation cost in the inference phase.

## 2.6. Multi-view Training, Inference, and Majority Voting

Previous studies mostly used only a single-view 2D slice data [20, 27], which lacks 3D structure information. Recent studies [23] started making use of multi-view data to leverage the extra information provided by multiple views. However, different 2D networks were trained on different views [13]. In our work, we trained the same model across different views, which forces the network to use the same weights to capture structures at different viewing-angles. We developed a multi-view training approach to train on different views, transversal, coronal, and sagittal planes, at a ratio of 4:1:1. Weight-sharing brings at least two benefits. First, compared to single-view training, training data is extended, so there will be less likely to be overfitting. Second, compared to multiple models for multi-view data, our single model requires fewer parameters and thus less memory. Each view provided a 3D segmentation prediction volume, and a simple majority voting strategy was applied to combine those three predictions into the final segmentation as shown in Figure 2. We used 3 GPUs in parallel for each subject. Thus, the inference time is determined by the slowest runtime among three views, which is just slightly longer than single-view inference.

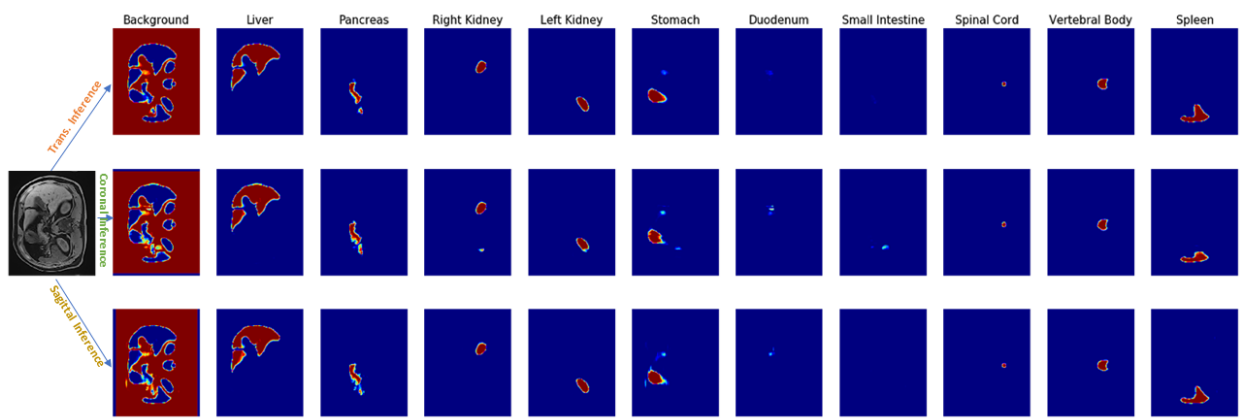

**Figure 2.** An example of the probability maps from three different view inferences. We fused three probability maps into one final segmentation result by the majority voting strategy.

**2.7. Evaluation Metrics**

Four metrics were used to evaluate the segmentation accuracy against the ground truth (i.e., human annotation). We chose two to indicate the ratio of volume overlapping: Dice Similarity Coefficient (DSC) [28] and Jaccard Index [29], and the other two to indicate the surface distance: mean surface distance (MSD) and 95% Hausdorff distance (95HD) [30].

**2.8. Network Implementation**

We implemented the network from ground-up in Tensorflow (v1.15) [31] deep learning packages. In this study, we trained the network with ADAM [32] optimizer. Its learning rate is 1e-4, and a decay rate of 0.9 was applied after every 50k iterations. We closely monitored the learning process by tracking the training and validation losses. We observed that models were well converged at 1000k steps (roughly 700 epochs) and used those checkpoints for evaluation. We used a workstation equipped with multiple Nvidia GTX 2080 TI Graphis Process Units (GPUs) for both training and testing.

# 3. RESULTS

**3.1. Optimization in the Network Size**

As shown in Table 1, when the network size increased, the runtime and parameter number increased in both DenseUnet and PlainUnet. In general, the PlainUnets were outperformed by DenseUnet by a notable margin, while the latter only used a fraction of the parameters thanks to its more efficient architecture. Furthermore, merely adding more parameters in PlainUnet did not significantly improve the performance. The DenseUnet ($k$=48) provided the best results in a reasonable runtime. Therefore, we use it as the baseline in our later experiments.

**3.2. The Effect of Different Normalization Methods**

We also examined the effect of different normalization methods, including a) no normalization, b) BN with the training mode in inference phase, c) BN with the testing mode in inference phase, d) LN, and e) IN. It turned out that no normalization provided the best results and fastest inference time, as shown in Table 2.

(a)

| | Liver | Pancreas | Right Kidney | Left Kidney | Stomach | Duodenum | Small Intestine | Spinal Cord | Vertebral Body | Spleen | Mean | Param# | Runtime |
|---|---|---|---|---|---|---|---|---|---|---|---|---|---|
| | **DenseUnet** | | | | | | | | | | | | |
| k24 | 0.956±0.013 | 0.791±0.082 | 0.947±0.006 | 0.941±0.013 | 0.868±0.048 | 0.686±0.083 | 0.780±0.135 | 0.873±0.028 | 0.834±0.037 | 0.931±0.030 | 0.861±0.084 | **1.55M** | **13.179** |
| k32 | 0.955±0.011 | 0.772±0.107 | 0.939±0.012 | 0.941±0.011 | 0.871±0.052 | 0.696±0.093 | 0.789±0.105 | 0.884±0.019 | 0.887±0.018 | 0.936±0.021 | 0.867±0.083 | 2.74M | 13.977 |
| **k48** | **0.960±0.009** | **0.828±0.074** | 0.940±0.009 | **0.951±0.008** | **0.889±0.046** | **0.732±0.076** | 0.790±0.103 | 0.866±0.028 | **0.889±0.017** | **0.934±0.018** | *0.878±0.071* | 6.12M | 18.052 |
| k64 | 0.959±0.012 | 0.799±0.086 | **0.952±0.008** | 0.950±0.009 | **0.889±0.046** | 0.708±0.091 | **0.796±0.107** | **0.886±0.024** | 0.882±0.026 | 0.929±0.025 | 0.875±0.079 | 10.83M | 21.040 |
| | **PlainUnet** | | | | | | | | | | | | |
| f64 | 0.957±0.012 | 0.740±0.131 | 0.943±0.010 | 0.943±0.012 | 0.870±0.045 | 0.695±0.075 | 0.792±0.098 | 0.888±0.024 | 0.884±0.016 | 0.908±0.037 | 0.862±0.086 | *229M* | *15.903* |
| f80 | 0.953±0.013 | 0.774±0.106 | 0.945±0.007 | 0.943±0.013 | 0.851±0.055 | 0.711±0.072 | 0.765±0.106 | 0.853±0.028 | 0.875±0.025 | 0.901±0.043 | 0.857±0.080 | 358M | 30.623 |
| f96 | 0.953±0.009 | 0.793±0.073 | 0.938±0.016 | 0.933±0.023 | 0.861±0.051 | 0.717±0.077 | 0.766±0.132 | 0.876±0.020 | 0.884±0.017 | 0.927±0.025 | 0.865±0.077 | 515M | 38.859 |

(b)

| | Liver | Pancreas | Right Kidney | Left Kidney | Stomach | Duodenum | Small Intestine | Spinal Cord | Vertebral Body | Spleen | Mean | Param# | Runtime |
|---|---|---|---|---|---|---|---|---|---|---|---|---|---|
| | **DenseUnet** | | | | | | | | | | | | |
| k24 | 1.329±0.433 | 2.750±1.319 | 0.800±0.181 | 0.973±0.578 | 3.326±1.546 | 4.315±1.457 | 4.177±3.421 | 0.795±0.143 | 1.655±0.359 | 1.273±0.842 | 2.139±1.314 | **1.55M** | **13.179** |
| k32 | 1.266±0.233 | 3.006±1.805 | 1.398±0.999 | 1.379±0.972 | 2.481±1.603 | 4.249±2.043 | 4.871±3.331 | 1.350±1.049 | 1.026±0.148 | 1.578±1.117 | 2.260±1.292 | 2.74M | 13.977 |
| **k48** | **1.188±0.269** | **2.270±1.167** | 1.212±0.893 | 0.934±0.493 | 2.732±1.241 | **2.931±1.295** | 5.081±3.932 | 0.875±0.219 | **1.097±0.200** | **1.180±0.632** | *1.950±1.269* | 6.12M | 18.052 |
| k64 | 1.218±0.367 | 3.324±2.849 | **0.738±0.165** | **0.743±0.205** | 2.367±0.976 | 4.275±1.832 | **4.604±3.441** | **0.738±0.110** | 1.461±1.523 | 1.433±0.978 | 2.090±1.405 | 10.83M | 21.040 |
| | **PlainUnet** | | | | | | | | | | | | |
| f64 | 1.318±0.367 | 2.664±1.395 | 1.212±0.796 | 1.121±0.669 | 2.799±1.100 | 4.352±2.010 | 4.760±2.888 | 0.811±0.264 | 1.089±0.219 | 1.666±1.042 | 2.179±1.346 | *229M* | *15.903* |
| f80 | 1.459±0.376 | 2.603±1.201 | 1.069±0.584 | 1.332±0.962 | 3.070±1.225 | 3.708±1.607 | 5.161±3.245 | 0.977±0.184 | 1.257±0.271 | 2.041±1.487 | 2.268±1.300 | 358M | 30.623 |
| f96 | 1.470±0.283 | 2.792±1.307 | 2.576±1.865 | 1.822±1.263 | 3.130±0.941 | 4.332±1.767 | 5.122±4.012 | 1.211±0.488 | 1.743±1.340 | 1.667±0.859 | 2.586±1.228 | 515M | 38.859 |

**Table 1.** (a) Dice Similarity Coefficient (DSC) and (b) Mean Surface Distance (MSD) between different network sizes of DenseUnet and PlainUnet on the test set. *K* is the growth rate of Densely Connected Block in DenseUnet, and *f* is the filter number of the first layer in PlainUnet. DenseUnet (*k*=32) runs faster than 100x larger PlainUnet(*f*=64) but still has moderately better performance, showing the advanced densely connections help in overall performance. DenseUnet (*k*=48) gains the best performance while still maintains a fast run time.

(a)

| | Liver | Pancreas | Right Kidney | Left Kidney | Stomach | Duodenum | Small Intestine | Spinal Cord | Vertebral Body | Spleen | Mean | Runtime |
|---|---|---|---|---|---|---|---|---|---|---|---|---|
| **NoNorm** | **0.961±0.008** | **0.860±0.042** | **0.954±0.006** | **0.952±0.009** | **0.907±0.024** | **0.766±0.066** | **0.839±0.085** | **0.898±0.021** | 0.886±0.015 | **0.944±0.013** | **0.897±0.059** | **8.307** |
| **BN_TrainMode** | 0.960±0.009 | 0.828±0.074* | 0.940±0.009* | 0.951±0.008 | 0.889±0.046* | 0.732±0.076* | 0.790±0.103* | 0.866±0.028* | 0.889±0.017 | 0.934±0.018* | *0.878±0.071** | *18.052* |
| **BN_TestMode** | 0.957±0.011* | 0.801±0.105* | 0.935±0.023* | 0.923±0.040* | 0.861±0.059* | 0.626±0.117* | 0.734±0.159* | 0.857±0.042* | 0.871±0.014* | 0.883±0.096* | *0.845±0.096** | *9.703* |
| **IN** | 0.960±0.009 | 0.826±0.068* | 0.944±0.007* | 0.948±0.010* | 0.888±0.042* | 0.726±0.078* | 0.782±0.116* | 0.851±0.039* | 0.874±0.028 | 0.935±0.023* | 0.874±0.074* | 13.216 |
| **LN** | 0.960±0.011 | 0.818±0.076* | 0.950±0.007* | 0.951±0.007 | 0.884±0.055* | 0.704±0.112* | 0.834±0.071 | 0.896±0.016 | **0.898±0.012*** | 0.940±0.013* | 0.883±0.076* | 13.503 |

(b)

| | Liver | Pancreas | Right Kidney | Left Kidney | Stomach | Duodenum | Small Intestine | Spinal Cord | Vertebral Body | Spleen | Mean | Runtime |
|---|---|---|---|---|---|---|---|---|---|---|---|---|
| **NoNorm** | **1.135±0.204** | **1.307±0.472** | **0.693±0.083** | 0.812±0.377 | **1.905±0.698** | **2.189±0.865** | **2.771±2.608** | **0.678±0.098** | 0.994±0.200 | **1.047±0.436** | **1.353±0.669** | **8.307** |
| **BN_TrainMode** | 1.188±0.269 | 2.270±1.167* | 1.212±0.893* | 0.934±0.493 | 2.732±1.241* | 2.931±1.295* | 5.081±3.932* | 0.875±0.219* | 1.097±0.200 | 1.180±0.632 | 1.950±1.269* | 18.052 |
| **BN_TestMode** | 1.370±0.290* | 2.265±1.153* | 1.006±0.449* | 1.058±0.572 | 3.420±2.691* | 4.192±2.462* | 3.890±2.526 | 0.933±0.254* | 1.211±0.340* | 2.129±2.216* | 2.147±1.196* | 9.703 |
| **IN** | 1.211±0.263 | 2.946±1.734* | 1.262±1.104* | 1.003±0.586 | 2.808±1.612* | 3.773±1.640* | 5.743±4.552* | 0.905±0.211* | 1.656±1.208* | 1.276±0.788 | 2.258±1.485* | 13.216 |
| **LN** | 1.179±0.282 | 1.726±0.841* | 0.714±0.138 | **0.808±0.218** | 2.078±0.882 | 2.653±1.144* | 3.476±3.081* | 0.691±0.140 | **0.922±0.142** | 1.259±0.740 | 1.550±0.886* | 13.503 |

**Table 2.** (a) Dice Similarity Coefficient (DSC) and (b) Mean Surface Distance (MSD) between different normalization methods (No Normalization, Batch Normalization [BN] in Train/Test mode, Instance Normalization [IN], and Layer Normalization [LN]) in DenseUnet ($k$=48) on the test set. * indicates significant difference (p < 0.05) compared to no normalization network. Without normalization, the network not only runs fastest but also obtains the best performance. Also, it is worth noting that using the training mode of BN during inference obtains a large performance improvement over using the testing mode of BN.

### 3.3. The Effect of Additional Training & Testing Techniques

To further improve the performance, we also tested the effects of additional training and testing techniques. We used 1) further augmentation with projective deformation (PD), 2) deep-supervised training (DS), 3) multi-view training (MTT), and 4) multi-view inference (MTI) with a majority voting in this experiment. The results are shown in Table 3. After incorporating all the above techniques, we observed a performance improvement over the baseline model.

(a)

| | Liver | Pancreas | Right Kidney | Left Kidney | Stomach | Duodenum | Small Intestine | Spinal Cord | Vertebral Body | Spleen | Mean | Runtime |
|---|---|---|---|---|---|---|---|---|---|---|---|---|
| Baseline | 0.961±0.008 | 0.860±0.042 | 0.954±0.006 | 0.952±0.009 | 0.907±0.024 | 0.766±0.066 | 0.839±0.085 | 0.898±0.021 | 0.886±0.015 | 0.944±0.013 | 0.897±0.059 | 8.307 |
| +PD | 0.962±0.008 | 0.864±0.034 | 0.953±0.007 | 0.952±0.007 | 0.913±0.027 | 0.775±0.055 | 0.846±0.091 | 0.898±0.024 | 0.895±0.016 | 0.944±0.011 | 0.900±0.056* | 8.370 |
| +PD+DS | 0.960±0.010 | 0.869±0.038* | 0.953±0.007 | 0.952±0.008 | 0.916±0.019* | 0.771±0.074 | 0.850±0.077* | 0.897±0.022 | 0.893±0.014 | 0.945±0.013 | 0.901±0.056* | 8.315 |
| +PD+DS +MTT | 0.961±0.011 | 0.870±0.042* | 0.954±0.007 | 0.954±0.009* | 0.913±0.022 | 0.782±0.069 | 0.860±0.063* | 0.895±0.017 | 0.895±0.011* | 0.945±0.014 | 0.903±0.053* | 8.327 |
| +PD+DS +MTT+MTI | 0.963±0.010 | 0.880±0.035* | 0.954±0.007 | 0.954±0.008* | 0.923±0.020* | 0.801±0.065* | 0.870±0.060* | 0.904±0.014 | 0.900±0.010* | 0.946±0.013 | **0.909±0.048*** | 12.25 |

(b)

| | Liver | Pancreas | Right Kidney | Left Kidney | Stomach | Duodenum | Small Intestine | Spinal Cord | Vertebral Body | Spleen | Mean | Runtime |
|---|---|---|---|---|---|---|---|---|---|---|---|---|
| Baseline | 1.135±0.204 | 1.307±0.472 | 0.693±0.083 | 0.812±0.377 | 1.905±0.698 | 2.189±0.865 | 2.771±2.608 | 0.678±0.098 | 0.994±0.200 | 1.047±0.436 | 1.353±0.669 | 8.307 |
| +PD | 1.117±0.248 | 1.215±0.303 | 0.685±0.118 | 0.693±0.137 | 1.736±0.590 | 2.128±0.759 | 2.593±2.942 | 0.663±0.102 | 0.938±0.196 | 1.063±0.645 | 1.283±0.627 | 8.370 |
| +PD+DS | 1.194±0.296 | 1.169±0.368* | 0.681±0.103 | 0.734±0.180 | 1.618±0.429* | 2.262±1.007 | 2.744±2.860 | 0.667±0.088 | 0.948±0.119 | 1.221±0.869 | 1.324±0.661 | 8.315 |
| +PD+DS +MTT | 1.137±0.303 | 1.167±0.363 | 0.657±0.084 | 0.655±0.094 | 1.618±0.505 | 2.254±0.924 | 1.959±1.709 | 0.675±0.076 | 0.939±0.106 | 1.027±0.505 | 1.209±0.532* | 8.327 |
| +PD+DS +MTT+MTI | 1.072±0.268 | 1.027±0.273* | 0.655±0.087 | 0.645±0.090* | 1.338±0.308* | 1.831±0.873* | 1.960±2.671 | 0.630±0.070* | 0.892±0.098* | 0.941±0.321 | **1.099±0.451*** | 12.25 |

**Table 3.** (a) Dice Similarity Coefficient (DSC) and (b) Mean Surface Distance (MSD) of different training and testing settings: projective deformation (PD), deep-supervised training (DS), multi-view training (MTT), and multi-view inference (MTI). * indicates significant difference ($p < 0.05$) compared to the baseline model: the non-normalized DenseUnet ($k$=48) network.

### 3.4. The Effect of Multi-Slice Training

Finally, we also investigated the results from the same DenseUnet structure with a different number of 2D slices as input to demonstrate the advantages of using multi-slice data. As shown in Table 4, a multi-slice input was beneficial for the overall performance compared with the single-slice input. However, the 40-slice version network was outperformed by the 20-slice version, suggesting that our current network structure could not process well with too many slices.

(a)

|  | Liver | Pancreas | Right Kidney | Left Kidney | Stomach | Duodenum | Small Intestine | Spinal Cord | Vertebral Body | Spleen | Mean |
|---|---|---|---|---|---|---|---|---|---|---|---|
| Single Slice | 0.963±0.012 | 0.871±0.038* | 0.953±0.007 | 0.955±0.008 | 0.920±0.021 | 0.771±0.051* | 0.870±0.066 | 0.902±0.017 | 0.898±0.014 | 0.941±0.019 | 0.904±0.055* |
| 20 Slices | 0.963±0.010 | 0.880±0.035 | 0.954±0.007 | 0.954±0.008 | 0.923±0.020 | 0.801±0.065 | 0.870±0.060 | 0.904±0.014 | 0.900±0.010 | 0.946±0.013 | **0.909±0.048** |
| 40 Slices | 0.964±0.009 | 0.871±0.041* | 0.953±0.008 | 0.955±0.008 | 0.921±0.021 | 0.788±0.074 | 0.873±0.062 | 0.904±0.018 | 0.899±0.011 | 0.943±0.022 | 0.907±0.051* |

(b)

|  | Liver | Pancreas | Right Kidney | Left Kidney | Stomach | Duodenum | Small Intestine | Spinal Cord | Vertebral Body | Spleen | Mean |
|---|---|---|---|---|---|---|---|---|---|---|---|
| Single Slice | 1.073±0.343 | 1.140±0.415 | 0.685±0.091* | 0.635±0.092* | 1.403±0.310 | 1.948±0.902 | 2.140±2.506 | 0.637±0.081 | 0.903±0.117 | 1.926±3.119 | 1.249±0.547* |
| 20 Slices | 1.072±0.268 | 1.027±0.273 | 0.655±0.087 | 0.645±0.090 | 1.338±0.308 | 1.831±0.873 | 1.960±2.671 | 0.630±0.070 | 0.892±0.098 | 0.941±0.321 | **1.099±0.451** |
| 40 Slices | 1.032±0.226* | 1.099±0.338* | 0.669±0.092* | 0.641±0.085 | 1.390±0.300 | 1.935±0.915 | 1.840±2.472 | 0.636±0.088 | 0.917±0.117* | 0.984±0.468 | 1.114±0.446 |

**Table 4.** (a) Dice Similarity Coefficient (DSC) and (b) Mean Surface Distance (MSD) of a different number of 2D stacked slice as input to the 2D DenseUnet. * indicates significant difference ($p < 0.05$) compared to our selected 20-slice setting. Multiple-slice input provided better performance than single slice input; however, adding more slices did not improve the results under the current network setting.

### 3.5. The performance of the finalized ALAMO method

Our finalized ALAMO method incorporated all the previous training & testing techniques in a non-normalized DenseUnet with 20-slice 2D data. The performance is summarized in Table 5. ALAMO system was able to achieve high-quality segmentation results on most of the organ with a DSC > 0.90, except for the duodenum (0.80) and the small intestine (0.87). A randomly selected test case is shown in Figure 3. We also show the box plot of per organ performance compared with different methods in Figure 4.

|  | Liver | Pancreas | Right Kidney | Left Kidney | Stomach | Duodenum | Small Intestine | Spinal Cord | Vertebral Body | Spleen | Mean |
|---|---|---|---|---|---|---|---|---|---|---|---|
| **DSC** | 0.963±0.010 | 0.880±0.035 | 0.954±0.007 | 0.954±0.008 | 0.923±0.020 | 0.801±0.065 | 0.870±0.060 | 0.904±0.014 | 0.900±0.010 | 0.946±0.013 | 0.909±0.048 |
| **Jacc** | 0.929±0.018 | 0.787±0.054 | 0.912±0.013 | 0.913±0.015 | 0.858±0.034 | 0.672±0.087 | 0.775±0.091 | 0.825±0.024 | 0.818±0.016 | 0.898±0.024 | 0.839±0.076 |
| **MSD** | 1.072±0.268 | 1.027±0.273 | 0.655±0.087 | 0.645±0.090 | 1.338±0.308 | 1.831±0.873 | 1.960±2.671 | 0.630±0.070 | 0.892±0.098 | 0.941±0.321 | 1.099±0.451 |
| **95HD** | 3.035±0.916 | 3.235±1.859 | 1.791±0.332 | 1.760±0.275 | 4.079±1.374 | 8.225±5.784 | 9.092±16.361 | 1.473±0.254 | 2.444±0.281 | 2.267±0.654 | 3.740±2.576 |

**Table 5**. Quantitative number of Dice Similarity Coefficient (DSC), Jaccard Index (Jacc), Mean Surface Distance (MSD), and 95 Hausdorff Distance (95HD) of our final ALAMO model on the test set (n=20).

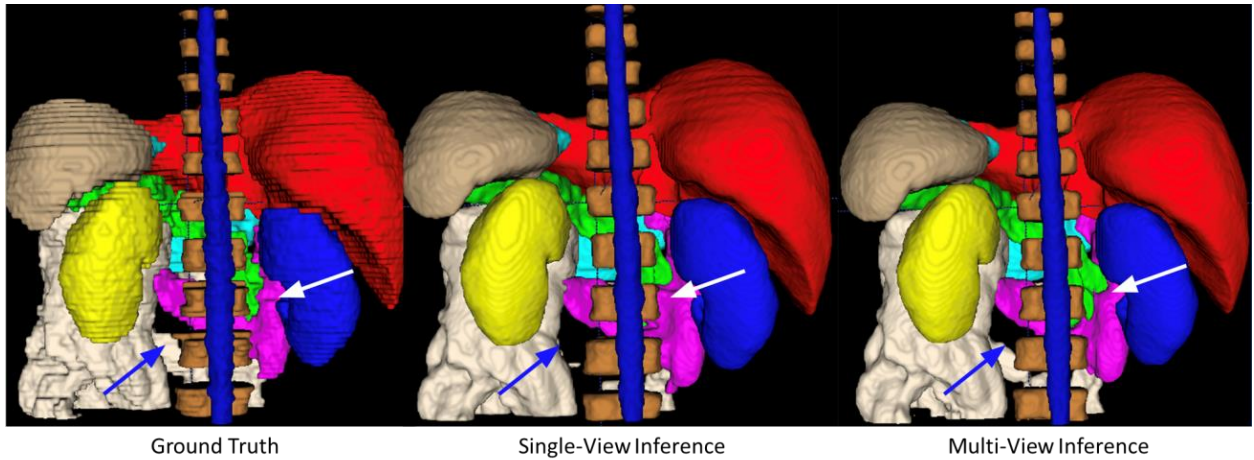

**Figure 3.** Segmentation results on a random test case with manual label, single-view (transversal) inference of DenseUnet, and multi-view inference with majority voting: liver(red), spleen(gray), pancreas(green), right kidney(blue), left kidney(yellow), stomach(cyan), duodenum(purple), small intestine(white), spinal cord (blue) and vertebral bodies (dark brown). The multi-view inference correctly segmented the small intestine that is missed in single-view inference, as shown by the blue arrow. Besides, it produced a more accurate boundary of the pancreas and duodenum, as shown by the white arrow.

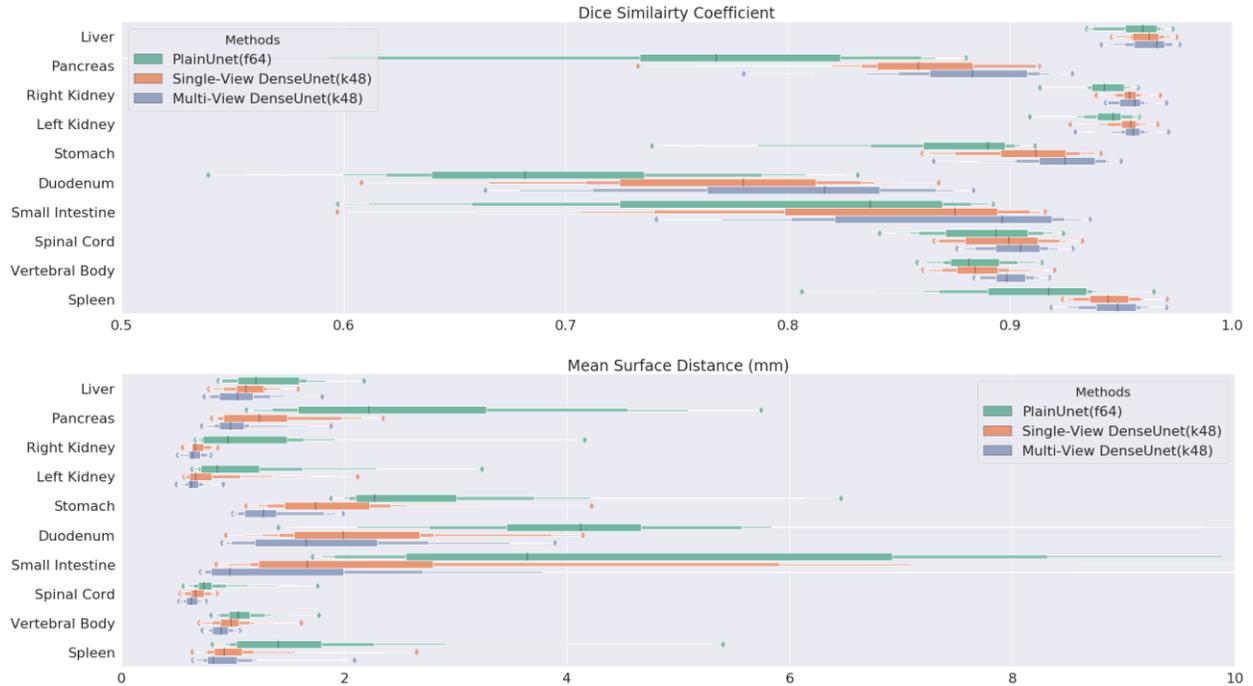

**Figure 4.** Per organ Dice similarity coefficient and mean surface distance box plot in test set (n=20) for PlainUnet(*f*=64), Single-View DenseUnet (*k*=48) and Multi-View Inference DenseUnet (*k*=48).

## 4. DISCUSSION

In this work, we presented a deep learning-based system ALAMO for fully automated multi-organ segmentation on abdominal MR. ALAMO builds on 2D DenseUnet and introduced tailored design in data augmentation and training procedures, utilizing deep connection, auxiliary supervision, and multi-view. The results showed that our system might be a strong candidate to afford state-of-the-art performance.

Given that medical imaging is intrinsically 3D, it is natural to consider a design with a 3D deep network as in [33, 34], which is associated with a much higher risk of overfitting and prohibitive demand in memory and computation time. Moreover, unlike computer vision tasks, the variations of the organ size and their relative geometric placement also make it necessary to have both high resolution and high spatial support in the network configuration. Even though it is possible to address such demand with a multi-resolution or hierarchical scheme, it introduces further complexity in the overall pipeline. We have demonstrated that a multi-slice-multi-scale 2D network, when carefully designed and optimized, provides a clinically viable alternative.

We compared two popular networks, PlainUnet and DenseUnet, in this paper. We showed that DenseUnet used much fewer parameters and offers more accurate segmentation results and slightly reduced computation time compared to PlainUnet. By adding multiple skip connections within the convolutional blocks, we force the network to reuse its weights, thus dramatically reduces the number of parameters for the same performance. Smaller network size not only makes training easier but also makes the model less prone to overfitting to the training data and more robust on unseen test data. This is critically important for MR-based DL applications due to typically limited data size.

The results confirmed our earlier speculation that normalization has insignificant performance gain in the single-sample batch setup. However, we want to point out that our observation does not exclude the possibility that a very thorough optimization of the normalization scheme may still yield some improvement. In our specific context, we feel the smaller computation burden and speed gain are more valuable for time-sensitive applications, and no-normalization would be the appropriate choice.

We also showed that combining three different views could further boost up performance. Forcing the network to train on different 2D views and then fusing them can remove some misclassified regions in single-view output. Especially for small intestine and duodenum, which have irregular shapes and are difficult to distinguish from complicated backgrounds, the multi-view network has a better performance compared to the single-view network. Additionally, since we used the same network in multi-view data, the model has fewer parameters than those where multiple models are used simultaneously. The reduction in parameters is beneficial to ease the overfitting problem.

Our last experiment also validated our assumption that by using multi-slice (both 20-slice and 40-slice) input data, a 2D network can perform better than a single-slice input. However, due to the current network design, using 40- slice as input did not improve the final accuracy, and the 20-slice input appeared to be a good fit for the DenseUnet ($k$=48).

There are not many studies on the multi-organ segmentation in abdominal MR with deep learning. To the best of our knowledge, Fu et al. [14] recently demonstrated their results on a ViewRay MR dataset with a DSC of 0.953 in the liver, 0.931 in kidneys, 0.850 in the stomach, 0.866 in bowels, and 0.655 in the duodenum. Bobo et al.[15] showed the segmentation on whole body T2-weighted MR, with a DSC of 0.913 in the liver, 0.730 in left kidney, 0.780 in the right kidney, 0.556 in the stomach, and 0.930 in the spleen. Our work demonstrated a better or similar DSC in those organs, especially in challenging organs such as the stomach and duodenum. Compared to multi-organ segmentation on CT images that have better signal to noise ratio and spatial resolution, our results are still very competitive. For example, Wang et al. [13] reported a DSC of 0.98 in the liver, 0.97 in the left kidney, 0.98 in the right kidney, 0.95 in the stomach, and 0.97 in the spleen. Our numbers were very close to them even though their resolution (0.5 mm) is much higher than ours (1.2 mm). Our segmentation of organs like duodenum (0.80) and small intestine (0.87) is even much better than their CT-base segmentation (0.75 in duodenum and 0.80 in the small intestine). For the most studied single organ - pancreas, our DSC 0.88 is still on par with recent state-of-the-art deep learning-based segmentation works [35-38].

There is still room to improve in our network, particularly for organs like the small intestine and duodenum. First, a 2D model might not be able to capture all image features because of some organs have highly complex shapes and inconsistent textures and boundary interfaces. A 3D network might have some advantages to segment them. However, limited by the computation resources, one must make the 3D network or the input 3D patch very small. Such limitation severely curbs 3D networks' performance. In the future, we might explore more computational efficient 3D networks to ease the hardware limitation. Second, in our current T1w water phase images, some boundaries between the organs are very difficult to distinguish even by experienced radiologists. Adding more contrast images such as T1-weighted fat phase and T2-weighted images might further improve the segmentation.

## 5. CONCLUSION

The proposed ALAMO framework allows for fully automated abdominal MR segmentation with high accuracy and low memory and computation time demands.

## 6. ACKNOWLEDGEMENT

This work is in part supported by NIH grant (NIH/NCI 1R21 CA234637)

# REFERENCES

1. Lenchik, L., et al., *Automated Segmentation of Tissues Using CT and MRI: A Systematic Review.* Academic radiology, 2019.
2. Whitfield, G.A., et al., *Automated delineation of radiotherapy volumes: are we going in the right direction?* The British journal of radiology, 2013. **86**(1021): p. 20110718-20110718.
3. Kohlberger, T., et al. *Automatic multi-organ segmentation using learning-based segmentation and level set optimization.* in *International Conference on Medical Image Computing and Computer-Assisted Intervention.* 2011. Springer.
4. Kang, D.J., *A fast and stable snake algorithm for medical images.* Pattern Recognition Letters, 1999. **20**(5): p. 507-512.
5. Duquette, A.A., et al., *3D segmentation of abdominal aorta from CT-scan and MR images.* Computerized Medical Imaging and Graphics, 2012. **36**(4): p. 294-303.
6. Iglesias, J.E. and M.R. Sabuncu, *Multi-atlas segmentation of biomedical images: a survey.* Medical image analysis, 2015. **24**(1): p. 205-219.
7. LeCun, Y., Y. Bengio, and G. Hinton, *Deep learning.* nature, 2015. **521**(7553): p. 436-444.
8. Litjens, G., et al., *A survey on deep learning in medical image analysis.* Medical image analysis, 2017. **42**: p. 60-88.
9. Akkus, Z., et al., *Deep learning for brain MRI segmentation: state of the art and future directions.* Journal of digital imaging, 2017. **30**(4): p. 449-459.
10. Avendi, M., A. Kheradvar, and H. Jafarkhani, *A combined deep-learning and deformable-model approach to fully automatic segmentation of the left ventricle in cardiac MRI.* Medical image analysis, 2016. **30**: p. 108-119.
11. Dalmış, M.U., et al., *Using deep learning to segment breast and fibroglandular tissue in MRI volumes.* Medical physics, 2017. **44**(2): p. 533-546.
12. Gibson, E., et al., *Automatic multi-organ segmentation on abdominal CT with dense v-networks.* IEEE transactions on medical imaging, 2018. **37**(8): p. 1822-1834.
13. Wang, Y., et al., *Abdominal multi-organ segmentation with organ-attention networks and statistical fusion.* Medical image analysis, 2019. **55**: p. 88-102.
14. Fu, Y., et al., *A novel MRI segmentation method using CNN-based correction network for MRI-guided adaptive radiotherapy.* Medical physics, 2018. **45**(11): p. 5129-5137.
15. Bobo, M.F., et al., *Fully Convolutional Neural Networks Improve Abdominal Organ Segmentation.* Proc SPIE Int Soc Opt Eng, 2018. **10574**.

16. Huang, G., et al. *Densely connected convolutional networks*. in *Proceedings of the IEEE conference on computer vision and pattern recognition*. 2017.
17. Dou, Q., et al., *3D deeply supervised network for automated segmentation of volumetric medical images.* Medical image analysis, 2017. **41**: p. 40-54.
18. Rofsky, N.M., et al., *Abdominal MR imaging with a volumetric interpolated breath-hold examination.* Radiology, 1999. **212**(3): p. 876-884.
19. Tang, Y., *Deep learning using linear support vector machines.* arXiv preprint arXiv:1306.0239, 2013.
20. Ronneberger, O., P. Fischer, and T. Brox. *U-net: Convolutional networks for biomedical image segmentation*. in *International Conference on Medical image computing and computer-assisted intervention*. 2015. Springer.
21. Çiçek, Ö., et al. *3D U-Net: learning dense volumetric segmentation from sparse annotation*. in *International conference on medical image computing and computer-assisted intervention*. 2016. Springer.
22. Jégou, S., et al. *The one hundred layers tiramisu: Fully convolutional densenets for semantic segmentation*. in *Proceedings of the IEEE Conference on Computer Vision and Pattern Recognition Workshops*. 2017.
23. Li, X., et al., *H-DenseUNet: hybrid densely connected UNet for liver and tumor segmentation from CT volumes.* IEEE transactions on medical imaging, 2018. **37**(12): p. 2663-2674.
24. Masters, D. and C. Luschi, *Revisiting small batch training for deep neural networks.* arXiv preprint arXiv:1804.07612, 2018.
25. Ulyanov, D., A. Vedaldi, and V. Lempitsky, *Instance normalization: The missing ingredient for fast stylization.* arXiv preprint arXiv:1607.08022, 2016.
26. Ba, J.L., J.R. Kiros, and G.E. Hinton, *Layer normalization.* arXiv preprint arXiv:1607.06450, 2016.
27. Roth, H.R., et al. *Deeporgan: Multi-level deep convolutional networks for automated pancreas segmentation*. in *International conference on medical image computing and computer-assisted intervention*. 2015. Springer.
28. Sorensen, T.A., *A method of establishing groups of equal amplitude in plant sociology based on similarity of species content and its application to analyses of the vegetation on Danish commons.* Biol. Skar., 1948. **5**: p. 1-34.
29. Jaccard, P., *Étude comparative de la distribution florale dans une portion des Alpes et des Jura.* Bull Soc Vaudoise Sci Nat, 1901. **37**: p. 547-579.
30. Huttenlocher, D.P., G.A. Klanderman, and W.J. Rucklidge, *Comparing images using the Hausdorff distance.* IEEE Transactions on pattern analysis and machine intelligence, 1993. **15**(9): p. 850-863.

31. Abadi, M., et al. *Tensorflow: A system for large-scale machine learning.* in *12th {USENIX} Symposium on Operating Systems Design and Implementation ({OSDI} 16).* 2016.
32. Kingma, D.P. and J. Ba, *Adam: A method for stochastic optimization.* arXiv preprint arXiv:1412.6980, 2014.
33. Chen, Y., et al. *Brain MRI super resolution using 3D deep densely connected neural networks.* in *2018 IEEE 15th International Symposium on Biomedical Imaging (ISBI 2018).* 2018. IEEE.
34. Chen, Y., et al. *Efficient and accurate MRI super-resolution using a generative adversarial network and 3D multi-Level densely connected network.* in *International Conference on Medical Image Computing and Computer-Assisted Intervention.* 2018. Springer.
35. Cai, J., et al. *Pancreas segmentation in MRI using graph-based decision fusion on convolutional neural networks.* in *International Conference on Medical Image Computing and Computer-Assisted Intervention.* 2016. Springer.
36. Cai, J., et al., *Improving deep pancreas segmentation in CT and MRI images via recurrent neural contextual learning and direct loss function.* arXiv preprint arXiv:1707.04912, 2017.
37. Asaturyan, H., et al. *Advancing Pancreas Segmentation in Multi-protocol MRI Volumes Using Hausdorff-Sine Loss Function.* in *International Workshop on Machine Learning in Medical Imaging.* 2019. Springer.
38. Gong, X., et al., *Computer-aided pancreas segmentation based on 3D GRE Dixon MRI: a feasibility study.* Acta radiologica open, 2019. **8**(3): p. 2058460119834690.